\documentclass[aps,prl,reprint,showpacs,byrevtex]{revtex4-2}
\let\oldhat\hat
\renewcommand{\hat}[1]{\oldhat{\mathbf{#1}}}
\usepackage{titlesec}
\usepackage{array}

\usepackage{times,mathptm}
\usepackage[dvips]{graphicx,color}

\begin{document}
\title{Surface-induced ferromagnetism and anomalous Hall transport at Zr$_2$S(001)}
\author{Shuyuan Liu$^1$, Yanwei Luo$^{1,2}$, Chongze Wang$^1$, Hyunsoo Jeon$^{1}$, Yu Jia$^3$, and Jun-Hyung Cho$^{1*}$}
\affiliation{$^1$Department of Physics and Research Institute for Natural Science, Hanyang University, 222 Wangsimni-ro, Seongdong-Ku, Seoul 04763, Republic of Korea \\
$^2$College of Science, Henan University of Technology, Zhengzhou 450001,  People's Republic of China
  \\
$^3$Key Laboratory for Special Functional Materials of the Ministry of Education, Henan University, Kaifeng 475004, People's Republic of China}
\date{\today}

\begin{abstract}
Two-dimensional layered electrides possessing anionic excess electrons in the interstitial spaces between cationic layers have attracted much attention due to their promising opportunities in both fundamental research and technological applications. Using first-principles calculations, we predict that the layered bulk electride Zr$_2$S is nonmagnetic with massive Dirac nodal-line states arising from Zr-4$d$ cationic and interlayer anionic electrons. However, the Zr$_2$S(001) surface increases the density of states at the Fermi level due to the surface potential, thereby inducing a ferromagnetic order at the outermost Zr layer via the Stoner instability. Consequently, the time-reversal symmetry breaking at the surface not only generates spin-polarized topological surface states with intricate helical spin textures but also hosts an intrinsic anomalous Hall effect originating from the Berry curvature generated by spin-orbit coupling. Our findings offer a playground to investigate the emergence of ferromagnetism and anomalous Hall transport at the surface of nonmagnetic topological electrides.
\end{abstract}
\maketitle


\section{I. INTRODUCTION}


Surface ferromagnetism has been a longstanding issue in condensed matter physics because of its importance from both the fundamental and technological points of view~\cite{review1,review2,review3}. The surface electronic structure often differs from the bulk one because surface atoms have the reduced coordination number due to their broken bonds. For example, the (001) surface of a 4$d$ transition metal Rh increases the density of states (DOS) at the Fermi level $E_F$ via narrowing its associated energy bands, thereby being vulnerable to a surface ferromagnetic instability~\cite{Rh surface1,Rh surface2}. Such surface-induced ferromagnetism in otherwise nonmagnetic bulk material can be explained by the Stoner model of itinerant magnetism: i.e., the Stoner criterion~\cite{Stoner criterion}, where the product of the exchange integral and the surface DOS at $E_F$ in the nonmagnetic state is greater than 1, is satisfied to induce ferromagnetism.

In the past decade, there have been intense research efforts to explore the connections between symmetries and topologies of condensed matter~\cite{efforts1,efforts2}. As a compelling example of symmetry-protected topological states, massless Dirac fermions with fourfold degenerate band crossings of two doubly degenerate bands are jointly protected by the time-reversal symmetry (TRS) $T$ and space inversion symmetry $P$ supplemented by additional crystalline symmetry such as glide mirror symmetry or screw rotation symmetry~\cite{symmetry-protected1,symmetry-protected2,symmetry-protected4}. However, when $T$ or $P$ symmetry is broken, Dirac fermions are transformed into Weyl fermions with twofold degenerate band crossings of two singly degenerate bands~\cite{symmetry-protected4,liangliang}. Such symmetry-protected topological states having linear band crossings at nodes or along one-dimensional lines or loops in momentum space can be classified into Dirac, Weyl, Dirac-nodal-line (DNL), or Weyl-nodal-line semimetal states~\cite{topological1,topological2,topological3}. These gapless topological states with the nontrivial topology of bulk bands host spin-polarized surface states with helical spin textures. However, without sufficient symmetry protection, the degeneracies at such band crossings are lifted to form hybridization gaps via the inclusion of spin-orbit coupling (SOC), which generate the Berry curvature around gapped crossings~\cite{Gd2C-shuyuan,Y2C-liangliang}. Here, we introduce a nonmagnetic bulk system that has massive DNL states with SOC-induced gap openings, but its surface exhibits a ferromagnetic order. The resulting TRS-breaking ferromagnetic surface hosts the emergence of an intrinsic anomalous Hall effect originating from the Berry curvature. By utilizing the spin degree of freedom generated only at surface, it is highly promising to realize anomalous transport phenomena in future spintronics technologies.

Recently, two-dimensional (2D) layered electrides $A_{2}B$ have aroused great interest for their exotic electronic properties such as low work function, high electron mobility, and spin polarization~\cite{Gd2C-shuyuan,Y2C-liangliang,Ca2N-Nature2013,shuyuan-jpcc,Gd2O-PRB2022}. In such $A_{2}B$ electrides consisting of a three-atom-thick building block of $A$-$B$-$A$ stacks [see Figs. 1(a) and 1(b)], anionic excess electrons reside in the interstitial spaces between positively charged $A$-$B$-$A$ cationic layers. The global structure search method has been used to predict various types of 2D layered electrides such as alkaline-earth nitrides~\cite{alkaline-earth nitrides}, rare-earth carbides~\cite{alkaline-earth nitrides,rare-earth carbides,rare-earth halides}, rare-earth pnictides~\cite{rare-earth pnictides}, rare-earth chalcogenides~\cite{rare-earth pnictides}, rare-earth halides~\cite{rare-earth halides,rare-earth pnictides}, and transition-metal monochalcogenides~\cite{transition-metal monochalcogenides}. To date, the experimentally synthesized 2D electrides are relatively few with (i) nonmagnetic Ca$_2$N~\cite{Ca2N-Nature2013}, Hf$_2$S~\cite{Hf2S-Sci. Adv.2020}, Zr$_2$S~\cite{transition-metal monochalcogenides}, Sr$_3$CrN$_3$~\cite{SrCrN3}, Sr$_8$P$_5$~\cite{SrP}, and Sr$_5$P$_3$~\cite{SrP}, (ii) paramagnetic Y$_2$C~\cite{Electride-Y2C2014,Y2CFM-JACS2017,2018Y2C}, and (iii) ferromagnetic Gd$_2$C~\cite{Gd2C-Nat.Commun.2020} and YCl~\cite{YCl-exp-2021}. Interestingly, recent DFT calculations~\cite{shuyuan-prb,Hf2Seprint} for the nonmagnetic Hf$_2$S electride predicted the existence of a ferromagnetic order at the Hf$_2$S(001) surface, where the local DOS arising from the surface Hf atoms increases at $E_F$ and thus induces a Stoner instability.

In this paper, we investigate the surface-induced ferromagnetism and its associated anomalous Hall effect in a nonmagnetic bulk electride Zr$_2$S using first-principles density-functional theory (DFT) calculations. We find that bulk Zr$_2$S has a topological band structure with massive DNLs consisting of hybridized Zr-4$d$ cationic and interlayer anionic states, which exhibit a high DOS distribution between $-$0.3 and $-$1.3 eV below $E_F$. However, such hybridized states associated with the topmost Zr layer at the Zr$_2$S(001) surface shift toward $E_F$ due to the surface potential. The resulting increase in the DOS at $E_F$ induces surface ferromagnetism via a Stoner instability. Furthermore, we demonstrate that the (001) surface of the nonmagnetic bulk electride Zr$_2$S not only possesses highly spin-polarized topological surface states with intricate helical spin textures but also exhibits a surface anomalous Hall effect originating from the Berry curvature. The present results offer a playground to investigate the intriguing interplay between electride materials, surface ferromagnetism, and anomalous transport phenomena.

\begin{figure}[h!t]
\includegraphics[width=8.5cm]{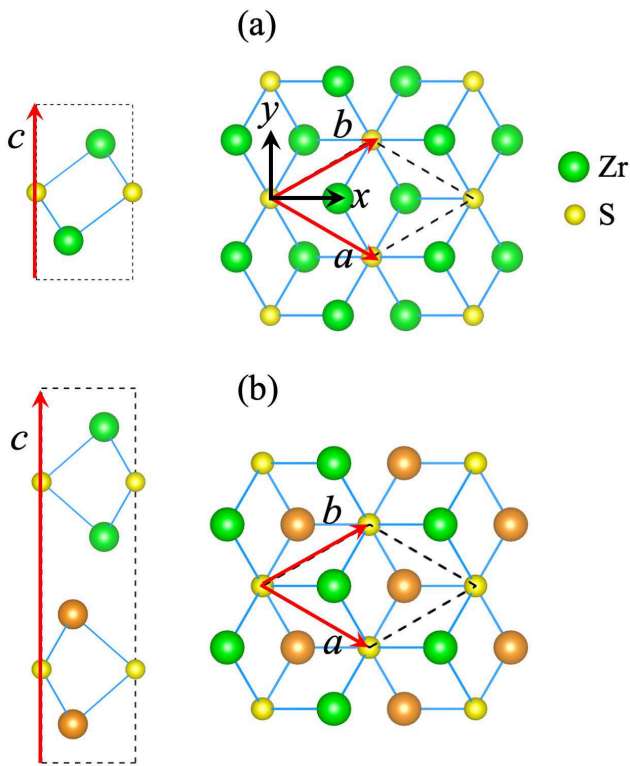}
\caption{Side (left panels) and top (right panels) views of the optimized structures of the (a) 1T and (b) 2H phases of bulk Zr$_2$S. The green and orange balls in panel (b) represent Zr atoms locating in neighboring layers. The lattice parameters $a$, $b$, and $c$ are drawn in each unit cell (indicated by the dashed lines). Here, the 1T phase has $a$ = $b$ = 3.548 {\AA} and $c$ = 5.600 {\AA}, while the 2H phase has $a$ = $b$ = 3.443 {\AA} and $c$ = 11.935 {\AA}.}
\label{figure:1}
\end{figure}

\section{II. CALCULATIONAL METHODS}

Our first-principles DFT calculations were performed using the Vienna $ab$ $initio$ simulation package (VASP) with the projector-augmented wave method~\cite{vasp1,vasp2,paw}. The exchange-correlation energy was treated with the generalized-gradient approximation functional of Perdew-Burke-Ernzerhof~\cite{pbe}. The plane wave basis was employed with a kinetic energy cutoff of 550 eV, and the $k$-space integration was done with 18${\times}$18${\times}$12, 18${\times}$18${\times}$6, and 18${\times}$18${\times}$1 meshes for the 1T bulk, 2H bulk, and (001) surface, respectively. All atoms were allowed to relax along the calculated forces until all the residual force components were less than 0.005 eV/{\AA}. The phonon spectrum calculation of the 1T bulk was carried out by using the QUANTUM ESPRESSO package~\cite{qe}, with the 6${\times}$6${\times}$4 $q$ points. The $ab$ $initio$ molecular dynamics simulations were performed by using a 3${\times}$3${\times}$2 supercell. The Zr$_2$S(001) surface was simulated using a periodic slab of twelve Zr-S-Zr stacks with ${\sim}$25 {\AA} vacuum in-between adjacent slabs.

\section{III. RESULTS AND DISCUSSION}

We begin by optimizing the 1T and 2H phases of bulk Zr$_2$S using DFT calculations. Here, the 1T phase crystallizes in an octahedral geometry with the space group $P\overline{3}m1$ (No. 164), while the 2H phase crystallizes in a trigonal prismatic geometry with the space group $P6_3/mmc$ (No. 194). Figures 1(a) and 1(b) show the optimized structures of the 1T and 2H phases, respectively. Our spin-polarized calculations for the two phases show that any initial ferromagnetic or antiferromagnetic configuration converges to a nonmagnetic one, indicating that bulk Zr$_2$S is nonmagnetic. We find that the 1T phase is more energetically favored over the 2H phase by 55.5 meV per 1T unit cell. Interestingly, the preferred 1T phase in Zr$_2$S contrasts with bulk Hf$_2$S where the 2H phase has been experimentally synthesized~\cite{Hf2S-Sci. Adv.2020}. Hereafter, we focus on the bulk and surface properties of the 1T phase of Zr$_2$S.

\begin{figure}[h!t]
\includegraphics[width=8.5cm]{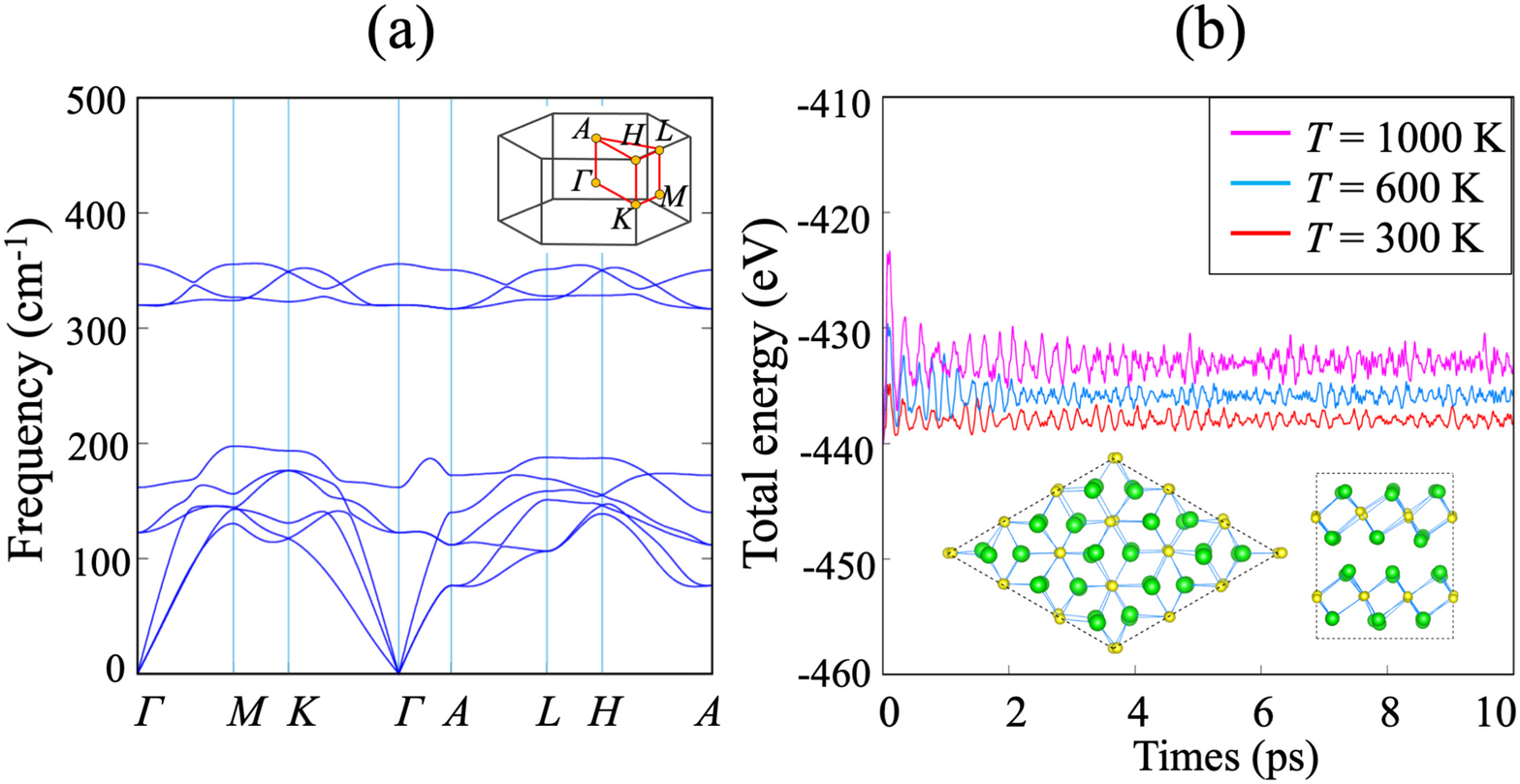}
\caption{(a) Calculated phonon spectrum of bulk Zr$_2$S. The total energy versus time for bulk Zr$_2$S, computed from $ab$ $initio$ molecular dynamics simulations at different temperatures, is displayed in panel (b). The inset in panel (b) shows the top and side views of the simulated structure at 1000 K after 10 ps.}
\label{figure:2}
\end{figure}

\begin{figure*}[h!t]
\includegraphics[width=17cm]{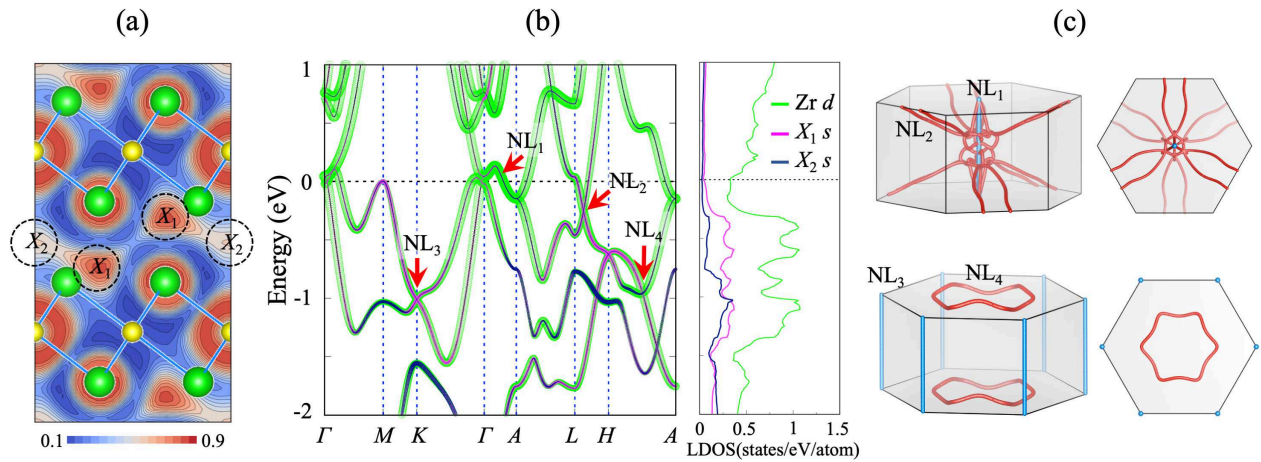}
\caption{(a) Calculated ELF of bulk Zr$_2$S on the (1$\overline{1}$0) plane with a contour spacing of 0.05. Here, the dashed circles represent $X_1$ and $X_2$ anions, the muffin-tin radii of which are chosen as 1 {\AA}. The calculated band structure and partial DOS of bulk Zr$_2$S are displayed in panel (b), where the projected bands on Zr-4$d$ and $X_1$-, $X_2$-$s$-like orbitals are represented by circles whose radii are proportional to the weights of the corresponding orbitals. In panel (b), the red arrows indicate various nodal lines NL$_1$, NL$_2$, NL$_3$, and NL$_4$ (see text) around $E_F$ whose dispersions over the Brillouin zone are drawn in panel (c). The right panels in (c) show the top views of nodal lines.}
\label{figure:3}
\end{figure*}

To examine the dynamic stability of bulk Zr$_2$S, we calculate its phonon spectrum using the density-functional perturbation theory~\cite{phonopy1,phonopy2}. As shown in Fig. 2(a), there are no imaginary phonon frequencies over the whole Brillouin zone, indicating that bulk Zr$_2$S is dynamically stable. We further perform $ab$ $initio$ molecular dynamics simulations to ensure the thermodynamic stability of bulk Zr$_2$S. Figure 2(b) shows the time evolution of total energies at different temperatures of 300, 600, and 1000 K. We find that the total energy at each temperature is well converged without large deviations. Specifically, the top and side views of the simulated structure at 1000 K after 10 ps show that the layered structure is preserved without any bond breakage. It is thus likely that bulk Zr$_2$S would be thermodynamically stable even at a high temperature of ${\sim}$1000 K. Recently, Zr$_2$S was experimentally synthesized with a polymorphic phase with the $P\overline{3}m1$ (equivalent to the 1T phase) and $Pnnm$ structures~\cite{transition-metal monochalcogenides}. Further experimental synthetic works will be demanded for the formation of a single 1T phase in the future.

Figure 3(a) shows the electron localization function (ELF) of bulk Zr$_2$S. It is seen that anionic excess electrons are well localized at the positions marked as $X_1$ and $X_2$ in the interlayer space, demonstrating that bulk Zr$_2$S is characterized as a 2D layered electride. In Fig. 3(b), the calculated band structure of bulk Zr$_2$S shows that Zr-4$d$ cationic and interstitial anionic states are strongly hybridized in the energy range between ${\approx}$$-$0.3 and ${\approx}$$-$1.3 eV below $E_F$, giving rise to several common peaks in their partial densities of states. We note that S-3$s$ and S-3$p$ orbitals are located below $-$3 eV (see Fig. S1 in the Supplemental Material~\cite{SM}), indicating that S atoms hardly participate in hybridization with interstitial anionic states. Interestingly, at the Zr$_2$S(001) surface, the hybridized Zr-4$d$ cationic and interstitial anionic states associated with the topmost Zr layer are shifted toward $E_F$, thereby inducing a surface ferromagnetic instability, as discussed below.

Using the tight-binding (TB) Hamiltonian with a basis of maximally localized Wannier functions~\cite{wannnier90,wanniertools}, we investigate the topological properties of bulk Zr$_2$S. The TB Wannier bands of bulk Zr$_2$S agree well with the corresponding DFT ones calculated using the VASP code (see Fig. S2 in the Supplemental Material~\cite{SM}). As shown in Figs. 3(b) and 3(c), we find fourfold degenerate band touching points around $E_F$, forming nodal lines NL$_1$ and NL$_3$ along the ${\Gamma}-A$ and $K-H$ paths, respectively, and nodal loops NL$_2$ and NL$_4$. It is noted that the NL$_2$ and NL$_4$ DNLs are respected by the $P$ symmetry, whereas the NL$_1$ (NL$_3$) DNL is respected by the $C_{3z}$ ($C_{3z}$ and $P$) symmetry~\cite{symmetry} (see Fig. S3 in the Supplemental Material~\cite{SM}). The nontrivial topological characterization of DNLs is identified by calculating the topological $Z_2$ index~\cite{Z2index}, defined as ${\zeta}_1$ = ${\frac{1}{\pi}}$ ${\oint}$$_c$ $dk$${\cdot}$A($k$), along a closed loop encircling each DNL. Here, A($k$) = $-i$$<$$u_k$$\mid$$\partial$$_k$$\mid$$u_k$$>$ is the Berry connection for the related Bloch states. We obtain ${\zeta}_1$ = ${\pm}$1 for the DNLs, indicating that they are stable against perturbations without breaking involved symmetries. However, by taking into account SOC, the fourfold degeneracy of DNLs is lifted to open gaps, thereby leading to massive DNLs around $E_F$.

Next, we investigate the electronic structure of the Zr$_2$S(001) surface. Figure 4(a) shows the band structure of the nonmagnetic  Zr$_2$S(001) surface in the absence of SOC. We find that there are two surface states $SS_1$ and $SS_2$ along the ${\Gamma}$-$K$-$M$ path near $E_F$, which are composed of the hybridized Zr-1 4$d$ cationic and interstitial $X_1$($X_2$)-$s$-like anionic states. Here, Zr-1, $X_1$, and $X_2$ represent Zr atom and interstitial anions at the topmost surface layer [see Fig. 4(b)].
It is noted that the number of electrons within the muffin-tin sphere of the surface $X_1$ ($X_2$) anion is 0.321 (0.204) electrons [see Fig. 4(b)], different from the corresponding bulk value of 0.570 (0.394) electrons [see Fig. 3(a)]. Figure 4(c) shows the local DOS (LDOS) projected on the Zr-1 and Zr-2 atoms [see Fig. 4(b)]. We find that the LDOS of Zr-1 at $E_F$ increases significantly compared to those of Zr-2 and bulk Zr [see Fig. 3(b)]. This dramatic change of Zr-4$d$ states together with the rearrangement of interstitial anionic electrons at Zr$_2$S(001) manifests strong surface effects. Due to the high LDOS of Zr-1 at $E_F$, the Stoner criterion may be satisfied to drive a ferromagnetic instability at the Zr$_2$S(001) surface. Indeed, the ferromagnetic phase is found to be energetically favored over the nonmagnetic one by 5.6 meV per Zr surface atom. For the ferromagnetic phase, the LDOS of Zr-1 shows that the spin-up and -down states are separated by ${\approx}$0.261 eV (see Fig. S4 in the Supplemental Material~\cite{SM}). By dividing this exchange splitting by the magnetic moment of Zr-1, we can estimate the Stoner parameter $I$, which in turn satisfies the Stoner criterion $I{\cdot}D(E_{F}) > 1$ (see Fig. S4~\cite{SM}). Here, $D(E_{F})$ is the LDOS of Zr-1 at $E_F$, obtained from the nonmagnetic surface. Thus, we can say that surface ferromagnetism emerging at the Zr$_2$S(001) surfaces is driven by the Stoner instability due to an increase in the surface DOS at $E_F$. To estimate the Curie temperature $T_{\rm c}$, we perform spin-polarized DFT calculations for various antiferromagnetic surface configurations (see Fig. S5~\cite{SM}). We find that the lowest antiferromagnetic configuration is less stable than the ferromagnetic one by 11.1 meV per Zr surface atom. Using the mean field approximation~\cite{MFA}, we estimate a $T_{\rm c}$ of ${\approx}$85 K at the Zr$_2$S(001) surface.

\begin{figure}[h!t]
\includegraphics[width=8.5cm]{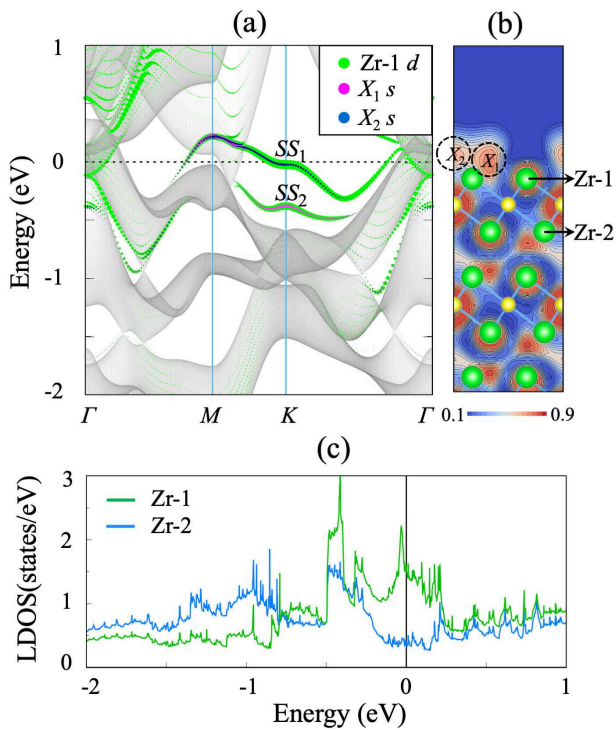}
\caption{(a) Calculated band structure of the nonmagnetic phase of the Zr$_2$S(001) surface. Here, the surface states $SS_1$ and $SS_2$ are projected onto the Zr-1 $d$, $X_1$ $s$-, and $X_2$ $s$-like orbitals where the radii of circles are proportional to the weights of the corresponding orbitals. The grey shaded region indicates the projection of bulk states. The ELF of the Zr$_2$S(001) surface with a contour spacing of 0.05 is displayed in panel (b), where the dashed circles represent $X_1$ and $X_2$ anions locating at the topmost surface layer. The calculated LDOS of Zr-1 and Zr-2 atoms at the Zr$_2$S(001) surface are given in panel (c).}
\label{figure:4}
\end{figure}

Figures 5(a) and 5(b) show the band structure and spin density of the ferromagnetic Zr$_2$S(001) surface, respectively, calculated without including SOC. We find that the $SS_1$ and $SS_2$ surface states are spin-polarized to exhibit surface ferromagnetism. Consequently, the spin density is mostly distributed around the Zr-1 layer, while it is significantly reduced at the Zr-2 layer [see Fig. 5(b)]. The calculated spin magnetic moments integrated within the muffin-tin spheres around Zr-1, $X_1$, and $X_2$ at the topmost surface layer are 0.189 ${\mu}_B$, 0.056${\mu}_B$, and 0.053${\mu}_B$, respectively (see Table I). It is noted that the magnitude of the spin magnetic moment of Zr-2 decreases to $-$0.014${\mu}_B$ and the Zr-1 and Zr-2 spins are antiferromagnetically coupled to each other. By including SOC, we calculate the magnetic anisotropy energy (MAE) at the Zr$_2$S(001) surface. Figure 5(c) displays the angular dependence of MAE on the $xy$-, $yz$-, and $zx$-planes. We find that the easy axis is out-of-plane with a MAE of 0.024 meV per Zr surface atom. As shown in Fig. 5(c), the MAE on the $xy$-plane is isotropic, whereas that on the $yz$- or $zx$-plane strongly depends on the angle ${\phi}$ relative to the $z$ direction. To explore the topological nature of surface states associated with the gapped bulk DNLs, we calculate their spin texture at $E_F$ with including SOC. Figure 5(d) shows the calculated Fermi surface at the ferromagnetic Zr$_2$S(001) surface. We find that the minority-spin $SS_1$ surface state forming a hole pocket around the $M$ point and an electron pocket along the ${\Gamma}-K$ path exhibits a helical spin texture with spin-momentum locking, indicating nontrivial topological surface states without backscattering~\cite{DNL-review,helicalspin}.

\begin{table}[ht]
\caption{Calculated spin magnetic moments (in unit of ${\mu}_B$ per Zr atom) of Zr-1 and Zr-2 at the Zr$_2$S(001) surface. The values of $X_1$ and $X_2$ at the topmost surface layer are also given.}
\begin{ruledtabular}
\begin{tabular}{lcccc}
   & Zr-1 &  Zr-2    &$X_1$  &$X_2$ \\  \hline
Magnetic moment  &0.189 &  -0.014 & 0.056 &  0.053 \\

\end{tabular}
\end{ruledtabular}
\end{table}

\begin{figure}[h!t]
\includegraphics[width=8.5cm]{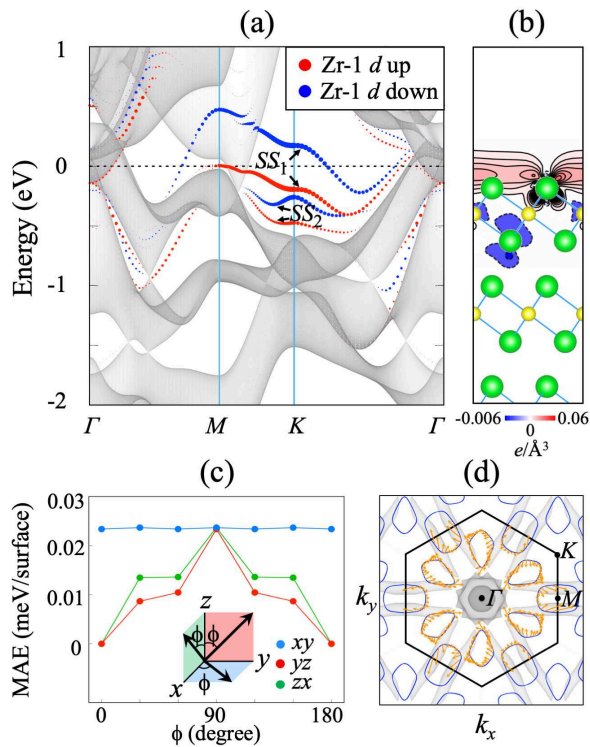}
\caption{(a) Calculated band structure of the ferromagnetic phase of the Zr$_2$S(001) surface. Here, the spin-polarized surface states $SS_1$ and $SS_2$ are projected onto the Zr-1 $d$ orbitals where the radii of circles are proportional to the weights of the corresponding orbitals. The grey shaded region indicates the projection of bulk states. In panel (b), the spin density at the Zr$_2$S(001) surface is drawn with a contour spacing of 0.005 $e$/{\AA}$^3$. The angular dependence of MAE with respect to ${\phi}$ on the $xy$-, $yz$-, and $zx$-planes is displayed in panel (c), where the minimum energy is set to zero. The spin texture of surface states on the Fermi surface is drawn in panel (d), where the arrows represent the $S_x$ and $S_y$ components along the horizontal and vertical directions, respectively. In panel (d), the blue lines represent the Fermi surface of the minority-spin $SS_1$ surface state, while the grey shaded region indicates the Fermi surface projected by bulk states.}
\label{figure:4}
\end{figure}

Finally, we examine the existence of an anomalous transport behavior at the ferromagnetic Zr$_2$S(001) surface. Using the Kubo-formula approach in the linear response scheme~\cite{kubo}, we calculate the intrinsic anomalous Hall conductivity ${\sigma}_{xy}$ originating from the Berry curvature. Figure 6(a) shows ${\sigma}_{xy}$ versus energy plot, obtained by integrating the $z$ component of Berry curvature (${\Omega}_z$) of all the occupied bulk and surface states over the surface Brillouin zone. We find that ${\sigma}_{xy}$ has a negative value of ${\approx}-$0.5 $e^{2}/h$ at $E_{\rm F}$, while it exhibits a large positive peak of ${\approx}$1.5 $e^{2}/h$ at $-$0.084 eV below $E_{\rm F}$. Thus, as hole doping shifts $E_{\rm F}$ to $-$0.084 eV (see Fig. S6 in the Supplemental Material~\cite{SM}), ${\sigma}_{xy}$ can increase significantly. Figures 6(b) and 6(c) display ${\Omega}_z$ distributions integrated in the two different energy ranges: one is the $R_{\rm I}$ region between $-$0.148 and $-$0.015 eV and the other is the $R_{\rm II}$ region between $-$0.015 eV and $E_F$. For the $R_{\rm I}$ region, there are the ``hotspots" of positive ${\Omega}_z$ arising from the electron pocket of the surface state along the ${\Gamma}-K$ path and the gapped NL$_1$ DNL around the ${\Gamma}$ point. Meanwhile, for the $R_{\rm II}$ region, the latter gapped DNL mostly contributes to more positive and negative ${\Omega}_z$ values. Therefore, the gapped NL$_1$  DNL plays an important role in ${\sigma}_{xy}$ around $E_F$, yielding an anomalous electronic transport behavior.

\begin{figure}[h!t]
\includegraphics[width=8.5cm]{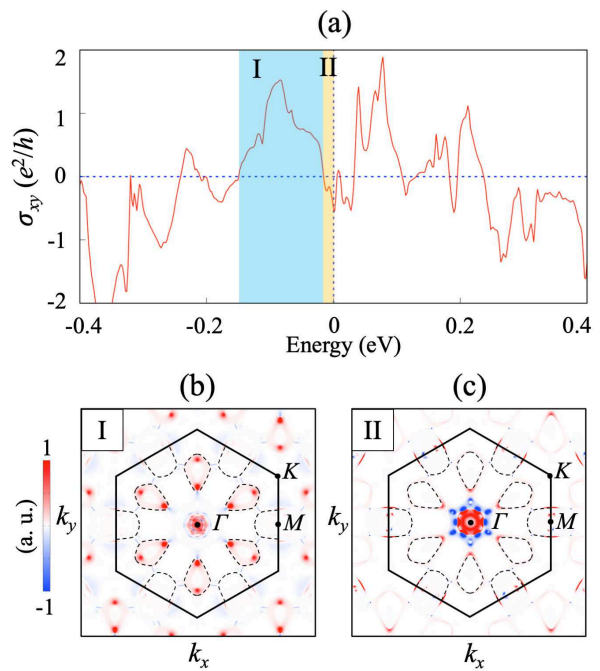}
\caption{(a) Energy dependence of ${\sigma}_{xy}$ and (b) (c) ${\Omega}_z$ distribution in the two energy regions I and II. In panels (b) and (c), red and blue areas indicate the ``hotspots" of positive and negative ${\Omega}_z$, respectively. The dashed lines in panels (b) and (c) represent the Fermi surface of the minority-spin $SS_1$ surface state.}
\label{figure:4}
\end{figure}

\section{IV. CONCLUSION}

Based on first-principles DFT calculations, we have predicted a ferromagnetic instability at the (001) surface of the 2D layered electride Zr$_2$S, the bulk of which is nonmagnetic. In contrast to bulk Zr$_2$S where hybridized Zr-4$d$ cationic and interlayer anionic states are located away below $E_F$, we found that the Zr$_2$S(001) surface has a high DOS at $E_F$ arising from such hybridized states, thereby inducing surface ferromagnetism via a Stoner instability. As a result, the Zr$_2$S(001) surface not only possesses highly spin-polarized topological surface states associated with massive DNLs but also hosts an intrinsic anomalous Hall effect originating from the Berry curvature generated by SOC. Therefore, our findings provide a novel platform to investigate the intriguing interplay between electride materials, surface ferromagnetism, and anomalous Hall transport which will be promising for future spintronics technologies.

\vspace{0.4cm}

\noindent {\bf Acknowledgements.}
This work was supported by the National Research Foundation of Korea (NRF) grant funded by the Korean Government (Grant No. 2022R1A2C1005456), by BrainLink program funded by the Ministry of Science and ICT through the National Research Foundation of Korea (Grant No. 2022H1D3A3A01077468), and by the National Natural Science Foundation of China (Grant No. 12074099). The calculations were performed by the KISTI Supercomputing Center through the Strategic Support Program (Program No. KSC-2022-CRE-0073) for the supercomputing application research.  \\


\noindent S. L. and Y. L. contributed equally to this work.  \\
\noindent $^{*}$ Corresponding author: chojh@hanyang.ac.kr

\end{document}